# Efficient Parallelization of Message Passing Neural Network Potentials for Large-scale Molecular Dynamics

Junfan Xia[1,2] and Bin Jiang[1,2*]

1. State Key Laboratory of Precision and Intelligent Chemistry, University of Science and Technology of China, Hefei, Anhui 230026, China
2. School of Chemistry and Materials Science, Department of Chemical Physics, University of Science and Technology of China, Hefei, Anhui 230026, China

*: corresponding author: bjiangch@ustc.edu.cn

## Abstract

Machine learning potentials have achieved great success in accelerating atomistic simulations. Many of them relying on atom-centered local descriptors are natural for parallelization. More recent message passing neural network (MPNN) models have demonstrated their superior accuracy and become increasingly popular. However, efficiently parallelizing MPNN models across multiple nodes remains challenging, limiting their practical applications in large-scale simulations. Here, we propose an efficient parallel algorithm for MPNN models, in which additional data communication is minimized among local atoms only in each MP layer without redundant computation, thus scaling linearly with the layer number. Integrated with our recursively embedded atom neural network model, this algorithm demonstrates excellent strong scaling and weak scaling behaviors in several benchmark systems. This approach enables massive molecular dynamics simulations on MPNN models as fast as on strictly local models for over 100 million atoms, vastly extending the applicability of the MPNN potential to an unprecedented scale. This general parallelization framework can empower various MPNN models to efficiently simulate very large and complex systems.

# 1. INTRODUCTION

Recent years have witnessed the revolutionary development of machine learning (ML) methods in representing potential energy surfaces (PESs) of complex systems based on *ab initio* calculations.[1-39] These machine-learned potentials (MLPs) have achieved much higher flexibility than empirical force fields with much lower computational costs than on-the-fly *ab initio* molecular dynamics (AIMD) simulations. One particularly successful family of MLPs is based on an atom-wise representation of the PES.[2] The total energy of the system is expressed as a sum of individual atomic energies, each of which is determined by many-body features within an atom-centered environment given by a cutoff radius ($r_c$). Consequently, such atom-wise MLPs exhibit linear scaling with system size, ensuring high scalability to extended systems. Their local feature-based nature also allows for straightforward parallelization.[40, 41] The energy evaluation can be efficiently distributed across multiple processors in an atom-wise fashion, eliminating the need for costly interatomic communication.

More recently, message passing neural network (MPNN) based approaches have become increasingly popular for learning PESs in an end-to-end manner.[8, 11, 15, 17-19, 22, 23] In this framework, the system is represented as an atomistic graph constructed by nodes (atoms) and edges (internode connections), in which atomic features are formed by iteratively propagating the geometric message from neighboring nodes via their edge features. After several message-passing (MP) iterations ($T$, often $2 \leqslant T \leqslant 6$), MPNNs effectively learn both local high-order many-body interactions and certain non-local information beyond the original $r_c$, thus generally achieving superior accuracy compared to conventional local MLP models.[17] Initial MPNN models propagated symmetry-invariant functions of interatomic distances or angles.[8, 11, 15] More advanced versions now incorporate equivariant features that include directional information, enabling them to capture orientation-dependent atomic interactions.[18, 21-23, 42-45] These developments greatly improve the structural discrimination, data efficiency, and transferability of MPNNs.

Despite these advantages, MPNN models face inherent limitations in parallelization scalability, which restricts their application to large-scale simulations of complex systems. This is because message communication occurs over $T$ sequential steps, effectively expanding the neighborhood radius to $Tr_c$. Consequently, all atoms within this effective $T$-dependent cutoff $r_{c,T}$ contribute to the final state of a central atom. It is commonly argued that the parallel computational cost of MPNNs scales unfavorably with $r_{c,T}$ or equivalently $T$, due to the cubically growing number of interacting atoms.[22, 46-48] To circumvent this difficulty, Kozinsky *et al*. proposed Allgero[47], a strictly local equivariant architecture that confines information exchange within the original cutoff radius. This simplifies its parallelization to be the same as other local MLPs. Alternatively, the MACE model[22] proposed by Csanyi *et al*. reduces $T$ to a maximum of two by utilizing high body-order equivariant messages to limit the parallel cost. Yutack *et al.* proposed SevenNet[48], a spatial decomposition algorithm based on the NequIP[21] architecture, which restricts the communication range between neighboring processors up to $r_c$ to limit redundant computations. However, this restriction prevents it from using an additional skin distance to reduce the frequency of updating neighbor lists, which could lower the efficiency of MD simulations in highly movable systems.

In this work, we propose a conceptually novel, general and efficient parallelization algorithm for MPNN models. A key concept is that the additional communication occurs only among local neighboring atoms in each MP layer, achieving a linear scaling rather than a cubic scaling with respect to $T$. This algorithm is implemented with a simple and physically inspired MPNN model, namely recursively embedded atom neural network (REANN) method[17]. Numerical tests on the various periodic systems validate that high efficiency of this algorithm, overcoming the common deficiency of all MPNNs. Our implementation also shows strong scaling with the number of CPU cores/GPUs and weak scaling with the number of atoms. This algorithm enables parallel molecular dynamics simulations of MPNNs up to ~100 million of atoms and offers a promising scheme for their applications in complicated systems.

# 2. RESULTS AND DISCUSSIONS

## 2.1 REANN architecture

As a simple MP adaptation of a local descriptor-based embedded atom neural network (EANN)[49] framework, the REANN model[17]—illustrated in Figure 1— serves as an ideal case study for comparative analysis of parallelization strategies of MPNN and strictly local MLP models. Specifically, we start by applying physically motivated embedded atom density (EAD) features to characterize local atomic environments.[49] An EAD feature of a central atom $i$ is defined by the square of the linear summation of orbital functions located at all neighboring atoms within $r_c$,

$$\rho_i = \sum_{l=0}^{L} \sum_{l_x,l_y,l_z}^{l_x+l_y+l_z=l} \frac{l!}{l_x!l_y!l_z!} \left[ \sum_{j\neq i}^{N_c} c_j G_j(\mathbf{r}_{ij}) \right]^2 \tag{1}$$

where $N_c$ is the number of atoms inside the cutoff sphere, $G_j(\mathbf{r}_{ij})$ is an orbital function of the $j$th neighbor atom obtained by the contraction of $N_\varphi$ primitive Gaussian-type orbitals, each of which is characterized by its center, width, and angular momenta ($l = l_x + l_y + l_z$ and $l \leq L$), and $c_j$ is the corresponding element-wise orbital coefficient. Varying these hyperparameters forms a vector of $N_\varphi(N_\varphi+1)(L+1)/2$ local EAD features, which encodes three-body interactions implicitly and serves as the input of atomic NNs outputting atomic energies in the EANN model. Noticing that $c_i$ itself should inherently rely on the neighborhood of atom $i$, it can be expressed by an atomic NN whose input is the corresponding EAD feature vector of atom $i$,

$$\mathbf{c}_i^t = M_i^{t-1}\left[\boldsymbol{\rho}_i^{t-1}(\mathbf{c}_j^{t-1}, \mathbf{G}_j)\right], \quad 1 \leq t \leq T-1 \tag{2}$$

while the EAD features can be recalculated once all $c_i$ are obtained. Consequently, $\boldsymbol{\rho}_i^t$ and $\mathbf{c}_i^t$ are iteratively updated via Eqs. (1) and (2) in an end-to-end MP manner, corresponding to node features and messages in the MPNN terminology, where $M_i^{t-1}$ represents the $i$th atomic NN module in the $t$th MP layer. After $T$ layers of MP, the last atomic NN layer outputs the atomic energy, whose sum yields the total energy $E$. In this sense, REANN is a three-body invariant feature-based MPNN model, which reduces to a strictly local EANN model when $T$=1.

## 2.2 Parallelization Algorithm

Let us first address the scaling of MPNN models with respect to the number of MP

layers, which was not clearly discussed in literature. It is well-known that the evaluation cost of local MLPs of condensed-phase systems scales cubically with respect to $r_c$ due to the cubic growth of neighbors in the cutoff sphere. Because the effective cutoff radius in MPNNs incrementally increases with the layer number of MP (see Figure 2(a)), it was argued that the feature evaluation cost would scale with $t^3 r_c^3$ in the *t*th MP layer[47], resulting in an actual accumulative scaling of $\sim \mathcal{O}(T^4)$ and significant demanding of communication overhead across multiple processes in parallel computation. In practice, however, the MP procedure is sequential and in each MP layer interatomic interactions remain confined to an atom-centered sphere within $r_c$. Critically, only the original neighbor list is required for each atom, and once computed, it can be repeatedly used in different MP layers. No need to maintain an extensive neighbor list as $T$ increases. As a result, the overall computational cost and communication overhead of the MPNN model can scale as favorably as with $\sim \mathcal{O}(T)$. This is the key prerequisite for efficient parallelization of MPNN models.

Figure 3(a) illustrates schematically the spatial decomposition scheme employed in the mainstream MD softwares[50, 51] for parallel simulations based on atomistic force fields. In each MD step, the simulation box is partitioned into non-overlapping subdomains, with each subdomain assigned to a unique process. Each process stores (i) positions/velocities and a complete set of neighbor lists of the local atoms (LAs), *e.g.*, atoms 1 and 2 (3 and 4) in process $P_1$ ($P_2$); (ii) positions/velocities for a group of ghost atoms (GAs), which lie within the cutoff sphere of its LAs but belong to adjacent subdomains, *e.g.*, atom 3 (2) in $P_1$ ($P_2$). Atoms are reassigned to new processes as they move through the physical domain. This design ensures each process to compute interactions which involve assigned LAs, but have accessibility to all necessary neighbors within the specified cutoff radius through GAs, enabling correct energy and force calculations across subdomain boundaries. Figure 3(b) shows clearly how this scheme works for the local EANN model via a computational graph, where downward and upward arrows represent forward (energy) and backward (force) workflows, respectively. Taking the forward inference as an example, all atomic coordinates (and coefficients) associated with GAs have been pre-assigned in each process in the neighbor list of LAs, enabling independent computation of atomic features of LAs and thus atomic energies.

However, as illustrated in Figure 3(c), nonlocal MPNN models cannot directly adopt this parallel algorithm. This is because orbital coefficients of GAs used for the feature generation of LAs in the current layer are virtually evaluated in the previous MP layer and distributed across different processes. They are messages that need to be passed from other processes rather than being directly calculated in the process where the neighbor lists of GAs are incomplete. For example, the incomplete neighbor list of the atom 3 as a GA in process $P_1$ would lead to a false feature vector, $\boldsymbol{\rho}_3^0$, and then cause continuous errors in relevant variables (marked in red). A similar issue arises in the backward evaluation for atomic forces, and further exacerbated by the chain broadcasting of erroneous tensor differentiation.

To address this challenge, we propose an efficient parallel algorithm for MPNN models. As illustrated in Figure 4(a), for the forward evaluation of the *t*th MP in each process, the messages of GAs are synchronized from the processes where they are classified as LAs and their features are correctly calculated. Meanwhile, the features of GAs in each MP layer are stored in relevant processes to keep the computational subgraph intact, enabling data communication and gradient calculations in the backward evaluation. As shown in Figure 4(b), atomic forces are calculated based on the chain rule of tensor differentiation in the backward evaluation. For atom *k*, the contributions to its force $\mathbf{F}_k$ can be evaluated in parallel within $N_p$ processes,

$$\mathbf{F}_k = -\sum_{p}^{N_p} \sum_{i \in L_p} \frac{\partial E_p}{\partial \boldsymbol{\rho}_i^{T-1}} \frac{\partial \boldsymbol{\rho}_i^{T-1}}{\partial \mathbf{r}_k}, \tag{3}$$

in which $L_p$ is the set of local atoms in process *p* and $E_p$ is the sum of their energies. Specifically, for the *t*th MP iteration in process *p*, we have,

$$\frac{\partial \boldsymbol{\rho}_i^t}{\partial \mathbf{r}_k} = \sum_{j \neq i}^{N_c} \left( \frac{\partial \boldsymbol{\rho}_i^t}{\partial \mathbf{c}_j^t} \frac{\partial \mathbf{c}_j^t}{\partial \mathbf{r}_k} + \frac{\partial \boldsymbol{\rho}_i^t}{\partial \mathbf{G}_j} \frac{\partial \mathbf{G}_j}{\partial \mathbf{r}_k} \right), \tag{4}$$

$$\frac{\partial \mathbf{c}_j^t}{\partial \mathbf{r}_k} = \frac{\partial \mathbf{c}_j^t}{\partial \boldsymbol{\rho}_j^{t\text{-}1}} \frac{\partial \boldsymbol{\rho}_j^{t\text{-}1}}{\partial \mathbf{r}_k}, \tag{5}$$

where $\frac{\partial \boldsymbol{\rho}_i^t}{\partial \mathbf{G}_j}$ and $\frac{\partial \mathbf{G}_j}{\partial \mathbf{r}_k}$ can be calculated locally in their respective processes, while only the gradient components associated with neighbor processes need data transfer, *e.g.*, $\frac{\partial \boldsymbol{\rho}_2^t}{\partial \mathbf{c}_3^t}$ and $\frac{\partial \mathbf{c}_2^t}{\partial \boldsymbol{\rho}_2^{t-1}}$ from $P_1$ to $P_2$ (Note that when *t*=0, $\frac{\partial \mathbf{c}_j^0}{\partial \mathbf{r}_k} = 0$). There is no need to synchronize the entire gradient tensor or the related computational subgraphs. In our implementation, both computational and communication overhead of gradients are further reduced by employing the vector-Jacobian product instead of explicitly computing the full Jacobian matrix. For example, in this way, the size of $\frac{\partial \boldsymbol{\rho}_2^t}{\partial \mathbf{c}_3^t}$ can be reduced by $N_\varphi(N_\varphi+1)(L+1)/2$ times. This parallel algorithm is done via the Message Passing Interface (MPI) based on the REANN architecture, referred to as REANN-MPI hereafter (see the Software section for details).

The proposed algorithm offers a general framework for parallelizing MPNN models. It is not limited to specific feature representations or MP schemes, requiring only the communication of GA-related data. In this sense, any MPNN models with messages aggregated over neighboring atoms, either based on invariant (*e.g.*, SchNet[8] and AIMNet[19]) or equivariant (*e.g.*, NequIP[21] and MACE[22]) features, can adopt a similar data communication scheme for their GA-related data in forward and backward evaluations. In comparison, the parallel scheme in SevenNet[48] is based on the communication of node features. For MPNN models where node features are updated by message pooling from neighboring atoms like REANN[17], SO3KRATES[27], or SEGNN[52], this scheme would incur repeated computation of GA message-related terms in both forward and backward passes across processes, resulting in noticeable increase of the parallelization overhead, even if the communication range is limited within $r_c$.

## 2.3 Performance of the algorithm

We first validate the proposed parallel algorithm by extensive MD simulations for solid Ag and liquid $H_2O$ systems, which have been frequently used to benchmark the performance of various MLP models[47, 53]. Figure 5 shows the relative cost of the REANN-MPI as a function of the number of MP layers for these two benchmark systems. To assess the actual computational cost, including communication overhead, the simulation box in each case is spatially decomposed into two equal-sized subdomains along each dimension, with each subdomain processed by a single CPU core. Figure 5 offers ambiguous numerical evidence that a properly parallelized MPNN model scales linearly with $T$ regardless of the system size or composition. The observed scaling behavior slightly deviates from the ideal relationship for systems comprising 1K atoms or fewer, where communication overhead becomes dominant. These findings overthrow the previous argument on the unfavorable scaling of $\sim\mathcal{O}(T^3)$ for most MPNN architectures[47] ($\sim\mathcal{O}(T^4)$ considering the accumulative cost).

Figure 6(a) demonstrates the strong scaling behavior of REANN-MPI, namely the computational speed *versus* the number of CPU cores (ranging from 1 to 1000) in MD simulations of liquid water (1.92 million atoms). Note that every CPU node in the test contains two Intel Xeon 9242 CPUs and 96 cores in total. Also shown are the results of the native Just-In-Time (JIT) implementation of REANN (REANN-JIT), which is a widely adopted parallel algorithm intrinsically supported in most Python-based platforms. It should be noted that REANN-JIT allows the inter-node MPI parallelization only for local models ($T$=1). In such a case, both REANN-MPI and REANN-JIT demonstrate comparable parallel performance, achieving near-linear speedups scaling to 1,000 CPU cores. This implies that our new parallel algorithm is compatible to local models while introducing negligible extra expense. When $T$>1, REANN-JIT is only compatible with the multi-threaded computation in a single node using Open Multi-Processing (OpenMP), which suffers from a very low parallel efficiency with many cores. By contrast, the REANN-MPI model retains a strong scalability and high efficiency, *e.g.* ~90% with 500 cores and ~80% with 1000 cores. This clearly demonstrates the superiority of the designed parallel algorithm. We note in passing that energies and forces evaluated by REANN-MPI and REANN-JIT are identical, validated by the same radial distribution functions of liquid water (see Figure S1 in Supplementary Materials).

Figure 6(b) further illustrates the scalability of the REANN-MPI model for both solid Ag and liquid $H_2O$ systems, where the number of nodes increases up to 32 while the number of atoms distributed in each node keeps fixed roughly from one million to three million. In each set of calculations, the actual speeds are nearly constant with the increasing number of nodes, demonstrating the excellent weak scaling behavior of the REANN-MPI model. Its absolute speed is also very fast, *e.g.*, reaching $7.4\times10^{-8}$ s/step/atom for over 100 million Ag atoms using 32 nodes with roughly 3 million atoms per node. This system size significantly surpasses the maximum scales previously achieved with non-local MPNN architectures[48, 54], matching the computational tractability of state-of-the-art strictly local MPNN models such as Allegro[47, 55]. In qualitative comparison with a well-recognized efficient local NN model, moreover, REANN-MPI is faster than the Deep Potential (DP) model under similar conditions in the simulation of liquid water[41], *e.g.*, $4.0\times10^{-7}$ s/step/atom for ~12.9 million of atoms *vs* $5.8\times10^{-7}$ s/step/atom for ~12.6 million of atoms, even though the latter utilizes ~4 times more CPU cores[41] (768 *vs* 3360). These results highlight the superior efficiency of our model and our parallel algorithm.

The new parallel algorithm naturally applies to GPU, since the implementation is based on LibTorch[56] which supports both CPU and GPU. Figure 7(a) compares the absolute speeds of the REANN-MPI and Allegro models[47] for roughly one million Ag atoms, with the number of GPUs increasing from 1 to 8. Note that the GPU node in the test equips 8 NVIDIA A100 GPUs, each with 80 GB memory. Impressively, our REANN-MPI model is in general two orders of magnitude faster the strictly local Allegro model, both exhibiting a similar strong scaling relationship, with the parallel efficiency slightly decreasing from ~88% to ~67% from 2 to 8 GPUs. The slight decline in efficiency with more GPUs may be attributed to the increasing cost of data transfer between the GPU and CPU, which can be migrated by direct data communication between GPUs in the future. Figure 7(b) demonstrates another qualitative comparison between REANN-MPI, SevenNet[48] and MACE-MP-0[54] in the simulation of a more challenging high entropy alloy (HEA) system consisting of five transition metal elements[57]. Interestingly, the REANN-MPI model, when loading 576 atoms/GPU, is approximately 5 times as fast as than SevenNet[48] in the same condition and 36~53 times than the MACE-MP-0 model[54] with 500 atoms/GPU, exhibiting an excellent weak scaling behavior from 1 to 8 GPUs. These results indicate that the proposed algorithm works well with GPUs and in more complex multi-element systems.

# 3. CONCLUSION

In summary, we propose an efficient algorithm for massive parallelization of MPNN potentials. A core concept of this algorithm is that the data communication is minimized among local neighboring atoms in each MP layer, warranting its superior efficiency. Numerical tests on our REANN model demonstrate that this algorithm allows a linear scaling with the MP layer instead of a power exponential scaling as previously argued for such models. The resultant REANN-MPI model exhibits strong scaling with the number of CPU cores and GPUs, as well as excellent weak scaling with the number of atoms across various benchmark systems. This approach enables fast parallel MD simulations involving over 100 million atoms, significantly extending the applicability of the MPNN potential. Our algorithm is not subject to a specific MP form and can be easily adapted for the parallelization of other MPNN models without changing their original structure. Further acceleration of the implementation of REANN-MPI can be achieved by enabling direct GPU-to-GPU data transfer. We anticipate that this algorithm will empower MPNN models to efficiently simulate very large and complicated systems that were only accessible by strictly local models.

# 4. METHODS

## 4.1 Systems and Training Details

Numerical tests of the proposed parallel algorithm have been done in several benchmark systems to validate its efficiency and generalizability. A single-element bulk crystal of Ag was used to demonstrate the favorable scaling of the REANN-MPI model with respect to system size, which can be directly compared with a strictly local MPNN model, Allegro, in that work with the same data set[47]. In Ref. 47, these DFT data of Ag were sampled from AIMD trajectories with the Vienna Ab-Initio Simulation Package (VASP) using the PBE exchange-correlation functional[58-60]. The dataset contains 1159 configurations consisting of 71 atoms, in which 1000 structures was used for training and validation, and 159 structures as a test set to calculate energy and force MAEs. The test MAEs of energy and force of the REANN model ($T$=2) are 0.255 meV/atom and 12.8 meV/Å, respectively, smaller than these values (0.397 meV/atom, 16.8 meV/Å) of the Allegro model reported in Ref. 47. To be consistent, MD simulations based on the REANN potential were also performed at a constant temperature of 300 K with a time step of 5 fs.

Bulk water is another important condensed phase system that has been frequently used to demonstrate the performance of MLPs. Here, we chose to fit the DFT dataset reported by Cheng *et al.*[53], consisting of energies and forces computed at the revPBE0-D3[61, 62] level for 1593 diverse structures of 64 water molecules in a cubic periodic box with a variable size. This dataset was randomly split into 80% for training set and 20% for test set as illustrated in Ref. 53. REANN model ($T$=2) yield smaller prediction RMSEs ($RMSE_E$=1.7 meV/atom, $RMSE_F$=71.2 meV/Å) than those reported in Ref. 53 ($RMSE_E$=2.3 meV/atom and $RMSE_F$=120.0 meV/Å). MD simulations based on the REANN potentials were performed at a constant temperature of 300 K with a time step of 0.2 fs.

High-entropy alloys (HEAs) are of great interest for their unique thermodynamic properties but with large chemical complexity, becoming a good candidate for testing the capability of the MLP and parallel algorithm in complex systems. To this end, we chose to fit the dataset reported by Lopanitsyna *et al.*[57], which contains 25,500 structures consisting of 3~25 transition metal elements while excluding Tc, Cd, Re, Os, and Hg. All energies and forces were computed using VASP with the PBESol functional[60, 63]. A total of 25,000 structures of this dataset were randomly selected for training, while the remaining structures were used for validation, as done in Ref. 57. The REANN model ($T$=2) yields smaller test MAEs ($MAE_E$=7.5 meV/atom and $MAE_F$=126.8 meV/Å) than those reported in Ref. 57 ($MAE_E$=10 meV/atom and $MAE_F$=190 meV/Å). MD simulations based on the REANN potentials were performed at a constant temperature of 300 K with a time step of 0.2 fs.

All relevant feature hyperparameters and NN structures of the REANN model in the three cases are listed in Table S1 of Supplementary Materials. It should be noted that the SevenNet model for the HEA system was trained with its public setting[48] as in https://github.com/MDIL-SNU/SevenNet.

### 4.2 Software Details

The new parallel algorithm is implemented in our REANN package, but rewritten in C++ with LibTorch, while the original REANN package is implemented in Pytortch equipped with the JIT technique. MD simulations were carried out using the LAMMPS code (10Feb2021). All the source codes were compiled with GCC (9.1.0), and linked to LibTorch (1.12.1), Intel MPI_CXX (3.1), OpenMP_CXX (4.5), and CUDA (11.3). Tests on SevenNet used newer versions of LibTorch (2.0.1) and LAMMPS (2Aug2023), as required by that model.

## Data availability

All datasets can be found in the corresponding original publications.

## Code availability

Code of REANN is available at https://github.com/bjiangch/REANN. The implementation of the MPI parallelization with REANN and the interface with LAMMPS is available at https://github.com/bjiangch/REANN-MPI.

## Conflicts of interest

The authors declare no competing financial interest.

## Acknowledgement

We thank the support by the Strategic Priority Research Program of the Chinese Academy of Sciences (XDB0450101), National Natural Science Foundation of China (22325304, 22221003 and 22033007). The simulations and model training were performed on the robotic AI-Scientist platform of Chinese Academy of Sciences, the Supercomputing Center of USTC, and Hefei Advanced Computing Center.

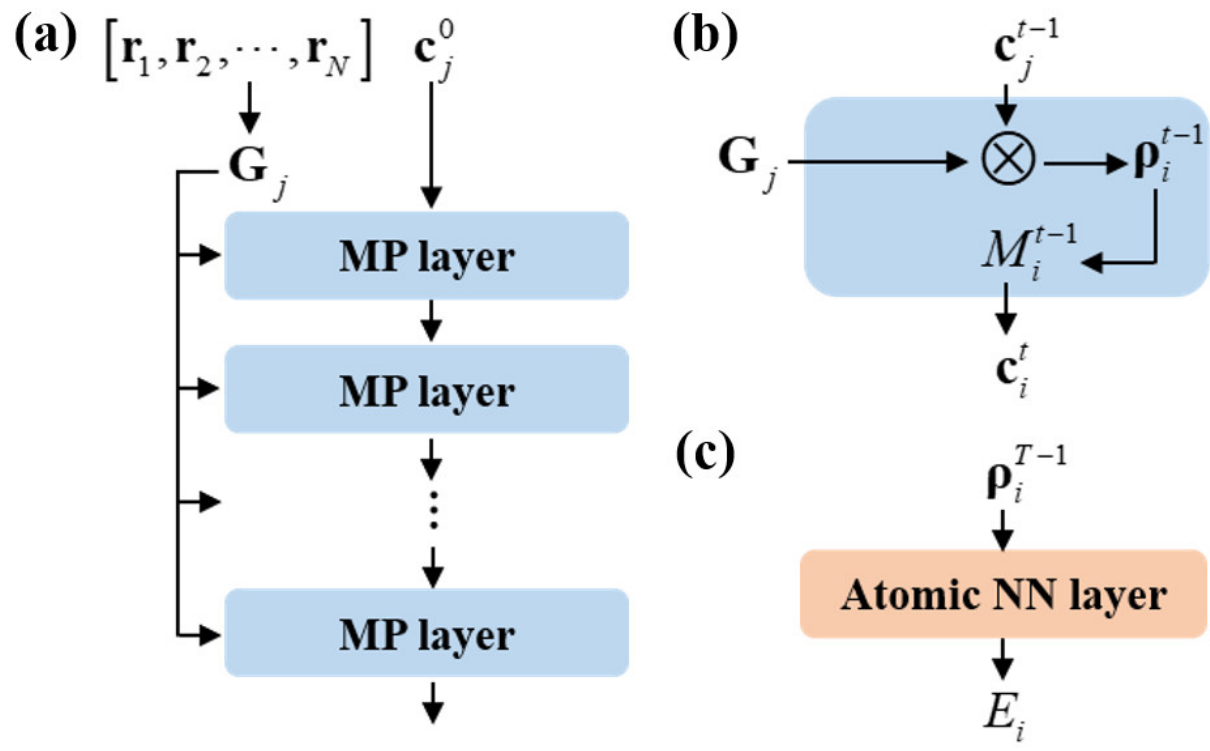

**Figure 1. Schematic diagram of the REANN framework. (a)** Input and MP layers, **(b)** each MP layer where $\otimes$ denotes the tensor product, and **(c)** the output layer after $T$ MP iterations.

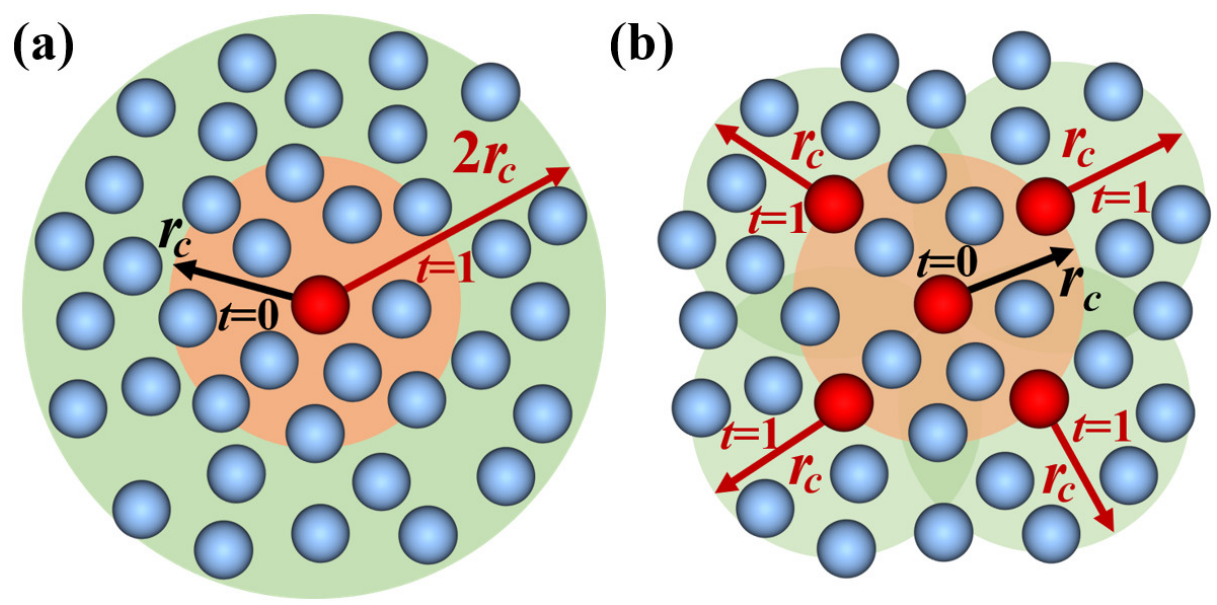

**Figure 2. Illustration of the growth of neighbor atoms via message passing. (a)** False impression considering all neighbor atoms of a fixed central atom as the effective cutoff radius increases. **(b)** Actual growth of the number of neighbor atoms in which each neighbor atom in the previous layer becomes a central atom of the current layer with the same cutoff radius. Supposing $r_c$ is the original cutoff radius and $T$=2.

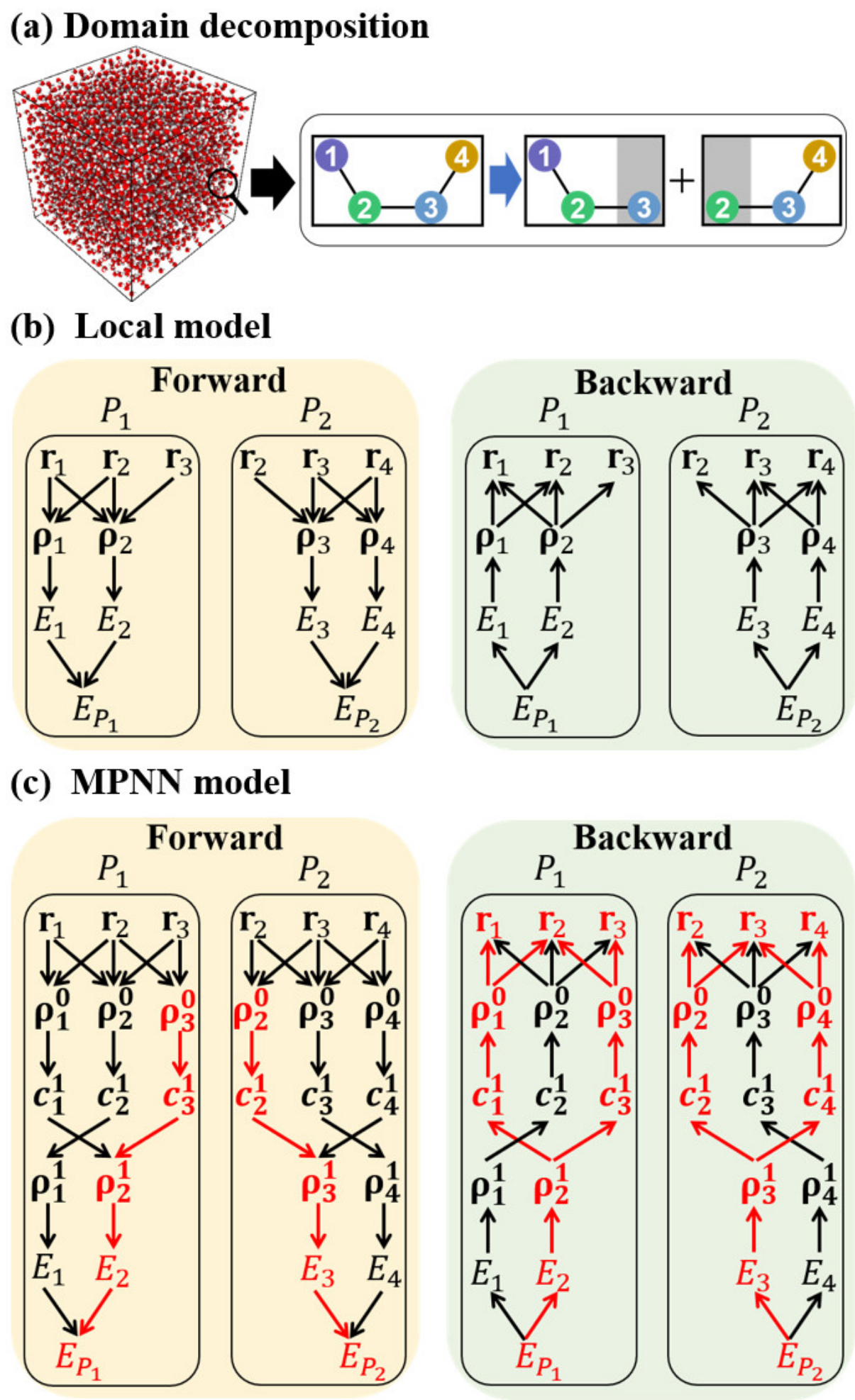

**Figure 3**. **Failure of the commonly-used parallel algorithm in MPNN models. (a)** Schematic diagram for domain decomposition in a toy system where local atoms and ghost atoms are respectively located in white and gray areas, with neighboring atoms connected. Computational graphs of **(b)** local models and **(c)** MPNN models ($T$=2) using the commonly-used parallel algorithm with two processes, $P_1$ and $P_2$. Forward (backward) evaluations between nodes (variables) are depicted by downward (upward) arrows. Nodes in the forward evaluation represent the calculation of variables themselves, while in the backward evaluation, they represent a set of derivatives of the previous nodes with respect to them. The red color indicates incorrect values due to the lack of the correct data transfer.

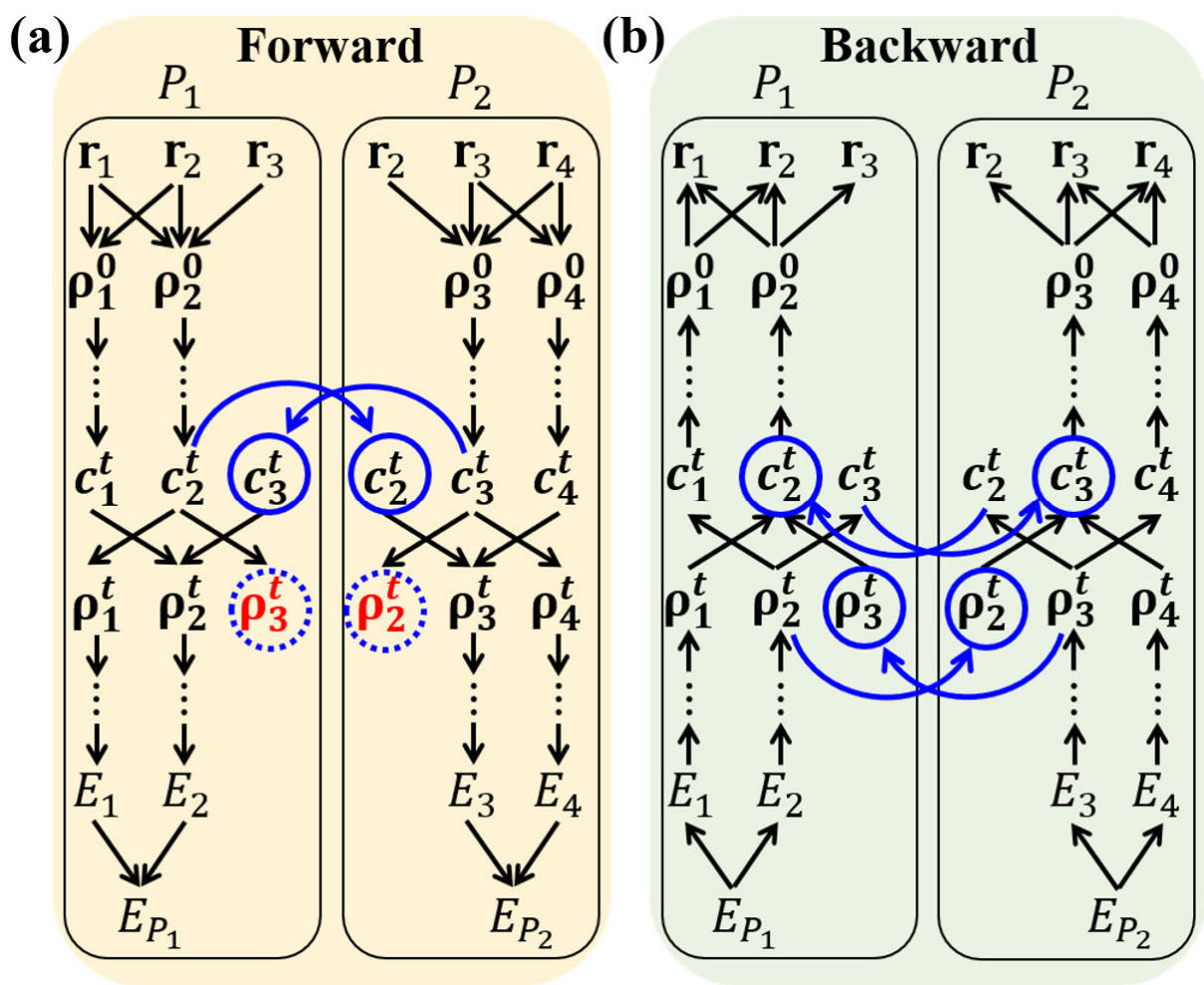

**Figure 4**. **Computational graphs of the proposed MPI parallel algorithm**. **(a)** Forward and **(b)** backward evaluations of MPNN models shown to, using with REANN as an example for illustration. Curved arrows indicate the inter-process data communications, solid circles denote variables that require data transfer, dotted circles indicate variables preserved for data communication of gradients in the backward evaluation of the computational subgraph.

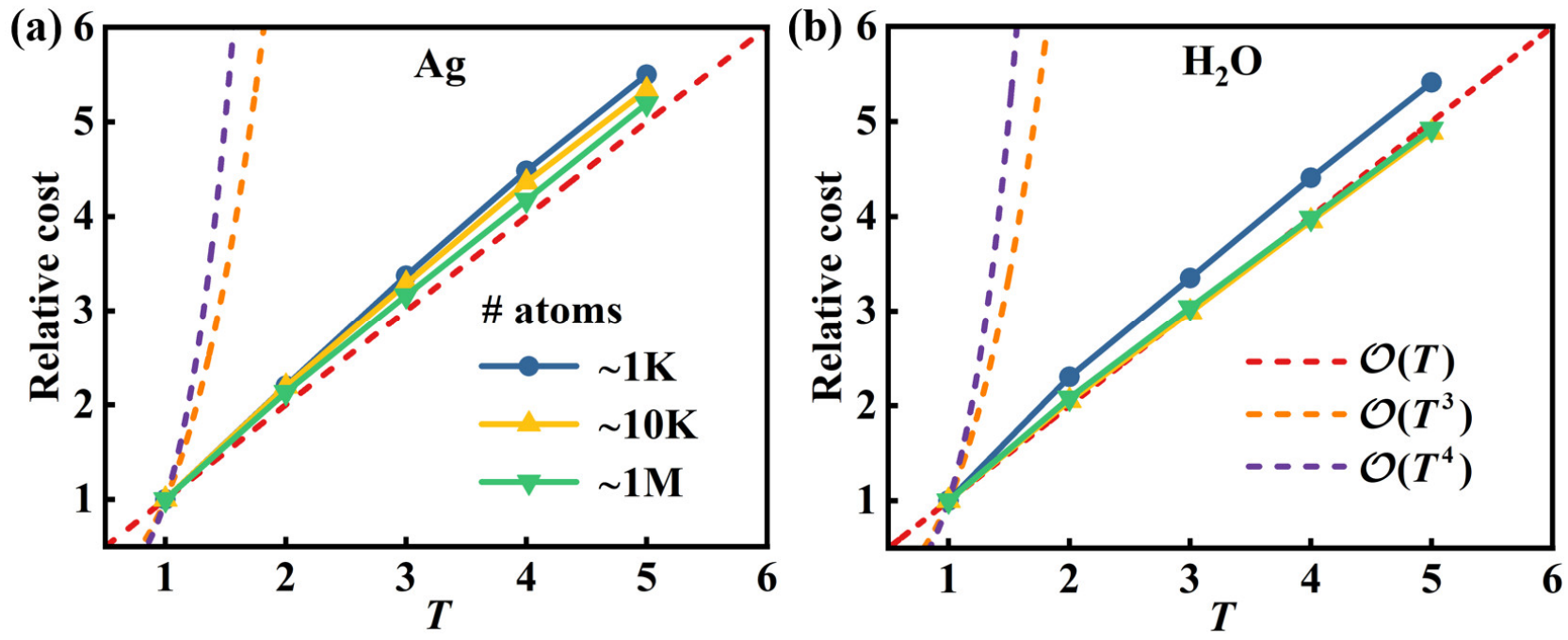

**Figure 5**. **Validating the model scaling with respect to message passing.** Comparison of relative costs of MD simulations for (a) Ag and (b) $H_2O$ using the REANN-MPI model, as a function of $T$ relative to the baseline cost of $T$=1. The number of atoms ranges from ~1 thousand (K) to ~1 million (M).

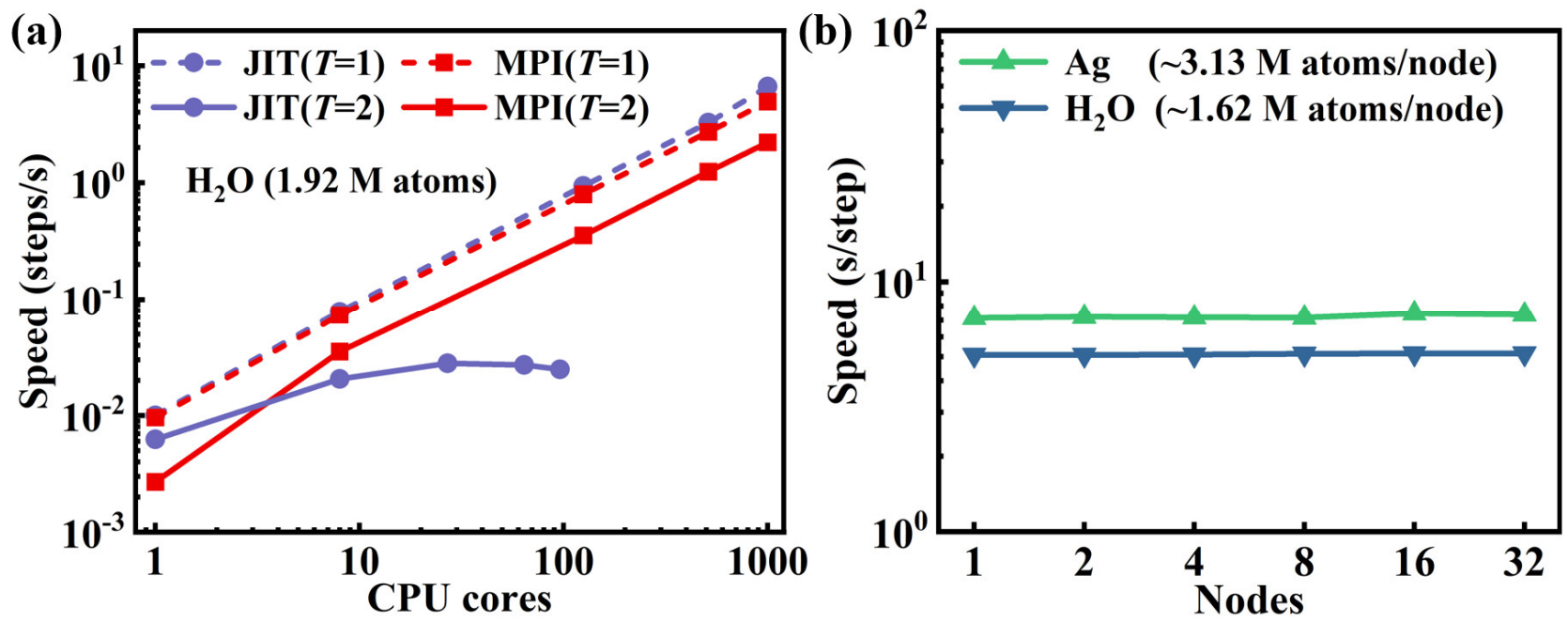

**Figure 6**. **Weak scaling and Strong scaling**. (a) Comparison of speed (*T*=1 and *T*=2) of REANN-MPI and REANN-JIT models as a function of CPU cores when simulating liquid water (1.92 million atoms). (b) Weak scaling performance of REANN-MPI in simulating both bulk Ag and liquid $H_2O$, where the number of atoms per node remains constant while the number of nodes increases from 1 to 32.

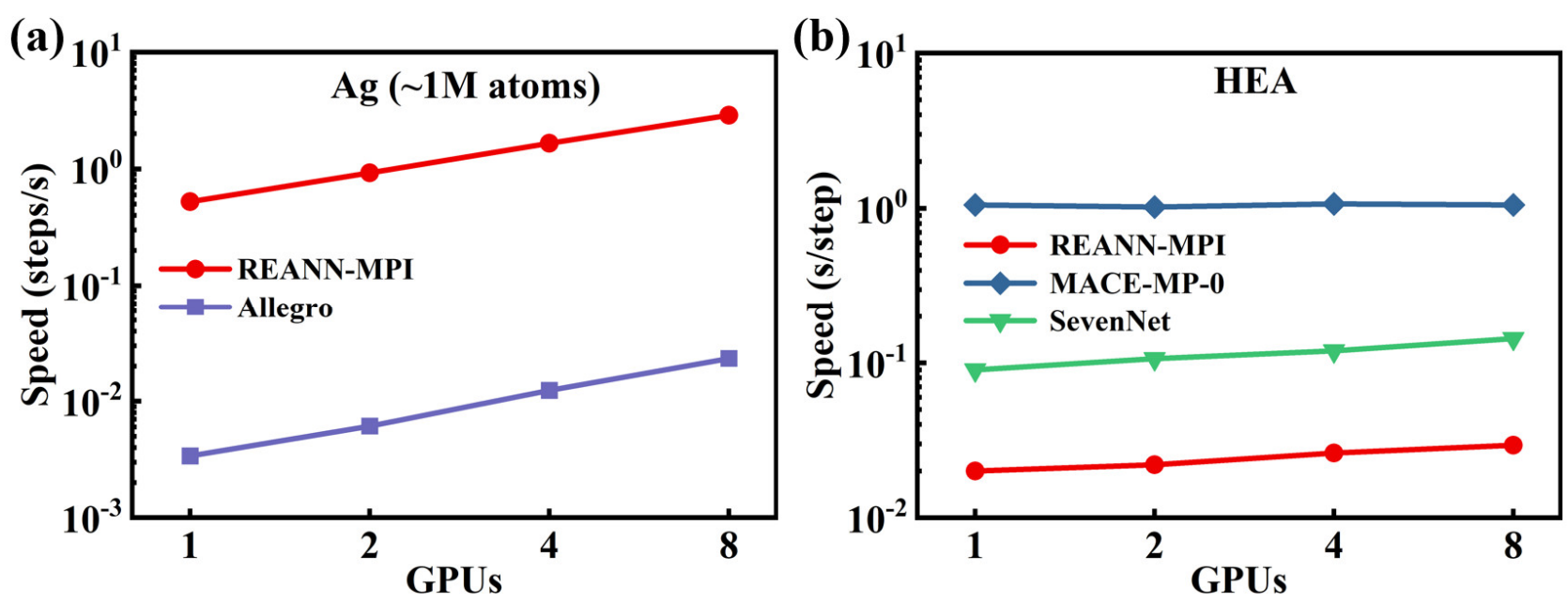

**Figure 7**. **Speed comparison with existing models.** (a) Speed comparison between REANN-MPI and Allegro in the simulations of Ag (1,022,400 atoms), (b) Speed comparison of SevenNet[48], MACE-MP-0[54], and the present REANN-MPI models in the simulation of HEA with 576 atoms/GPU, 500 atoms/GPU and 576 atoms/GPU, respectively, as a function of the number of NVIDIA A100 GPUs. The absolute cost data of the Allegro and MACE-MP-0 models were taken from the Ref. 47 and Ref. 54, respectively.

# Supplementary Materials

## Efficient Parallelization of Message Passing Neural Network Potentials for Large-scale Molecular Dynamics

Junfan Xia[1,2] and Bin Jiang[1,2*]

1. State Key Laboratory of Precision and Intelligent Chemistry, University of Science and Technology of China, Hefei, Anhui 230026, China
2. School of Chemistry and Materials Science, Department of Chemical Physics, University of Science and Technology of China, Hefei, Anhui 230026, China

*: corresponding author: bjiangch@ustc.edu.cn

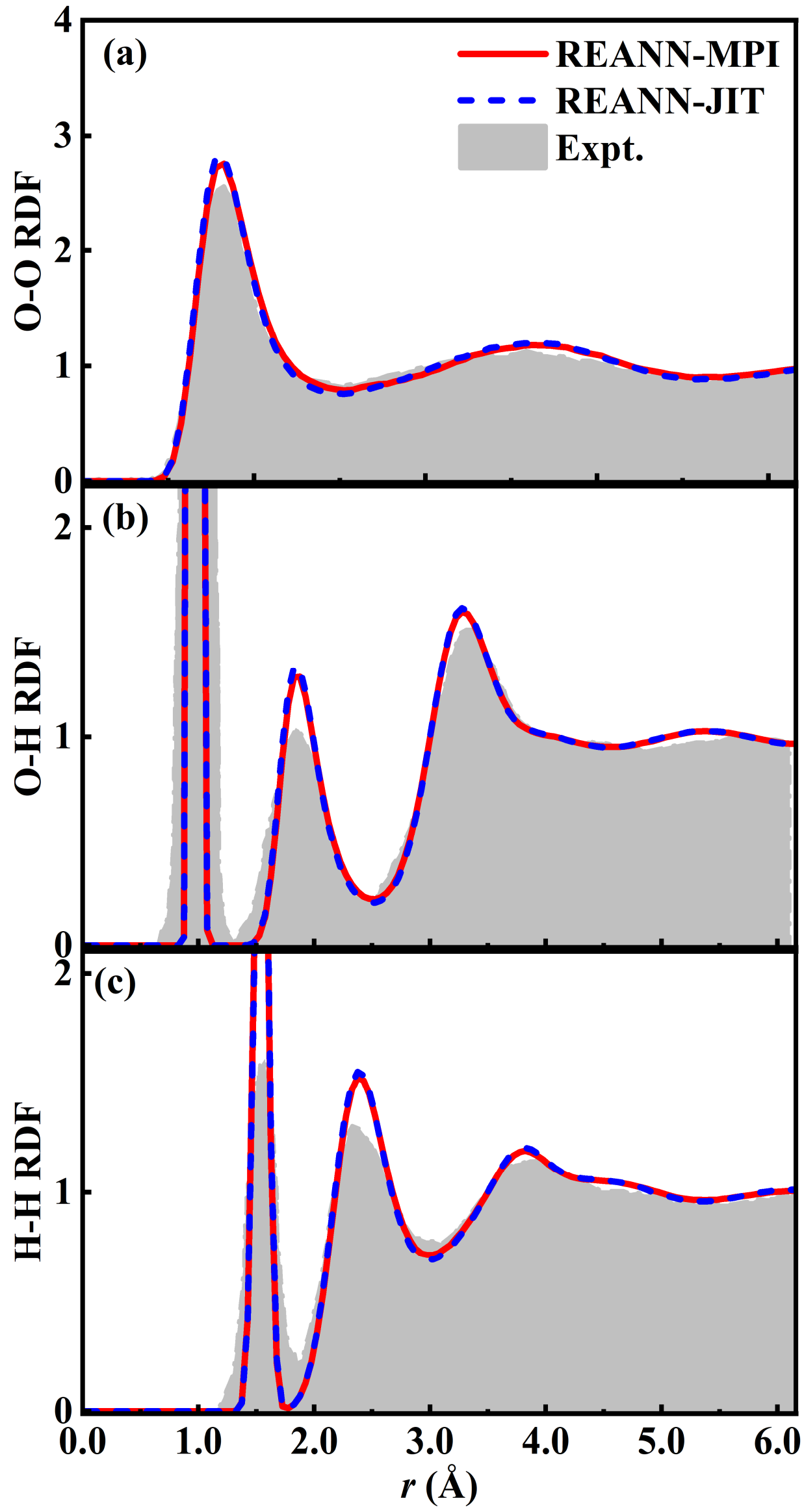

**Figure S1.** (a) O–O, (b) O–H, (c) and H–H RDFs obtained from the REANN-MPI and REANN-JIT models, and experimental data[1-3].

**Table S1**. Feature hyperparameters and NN structures (the number of neurons in each hidden layer) used in the training processes.

| **System** | ***L*** | $N_\varphi$ | $r_c$ (Å) | **Atomic NN layer structure** | **MPNN structure** |
|---|---|---|---|---|---|
| Ag | 2 | 7 | 4.0 | 32×16 | 16×16 |
| $H_2O$ | 2 | 7 | 3.8 | 32×16 | 16×16 |
| HEA | 2 | 7 | 6.0 | 32×16 | 16×16 |